\documentstyle[prl,aps,epsf]{revtex}
\def\<{\langle}
\def\>{\rangle}
\def\d{\partial}
\def\+{\dagger}

\def\U1A{U(1)$_{\rm A}$}
\def\LambdaQCD{\Lambda_{\rm QCD}}

\firstfigfalse

\begin{document}
\twocolumn[\hsize\textwidth\columnwidth\hsize\csname@twocolumnfalse\endcsname
\preprint{CU-TP-xxx}
\title{Domain Walls of High-Density QCD}
\author{D.T.~Son,$^{1,4}$ M.A.~Stephanov,$^{2,4}$ and A.R.~Zhitnitsky$^3$}
\address{$^1$Physics Department, Columbia University, New York, New York 10027}
\address{$^2$Department of Physics, University of Illinois, Chicago, Illinois
60607-7059}
\address{$^3$Department of Physics and Astronomy, University of
British Columbia, Vancouver, British Columbia, Canada V6T 1Z1}
\address{$^4$RIKEN-BNL Research Center, Brookhaven National Laboratory,
Upton, New York 11973}
\maketitle
\begin{abstract}

We show that in very dense quark matter there must exist
metastable domain walls where the axial U(1) phase of the 
color-superconducting condensate changes by $2\pi$.  The decay
rate of the domain walls is exponentially suppressed and we 
compute it semiclassically.  We give an estimate of the critical
chemical potential 
above which our analysis is under theoretical control.

\end{abstract}
\vskip 2pc]

{\em Introduction.---}%
Domain walls are common in field theory \cite{Rajaraman}.  They are
configurations of fields interpolating between two vacua.  If these
two vacua are distinct the domain wall cannot decay. There are, however,
theories with only a single vacuum, in which, nevertheless, 
domain-wall configurations exist.
The most notable example is the theory of $N=1$ axion \cite{axions}.
In such theories the decay of the domain wall is possible, though,
the decay rate is often suppressed.
It is generally believed that there are no domain walls of either kind
in the Standard Model. 
It was advocated only recently that long-lived 
domain walls may exist in QCD \cite{Zhitnitsky} at zero
temperature and baryon density. The possibility of
very unstable walls was noticed earlier in Ref.\ \cite{GabadadzeShifman}.
Unfortunately, no theoretical control
is possible in this nonperturbative regime.  
Somewhat related but different domain-wall configurations were also
discussed in Ref.\ \cite{KPT} in the context of the decay of the metastable
vacua possibly created in heavy ion collisions.

In this paper we show that, in the regime of high baryon densities, 
where relevant physics is under theoretical control, 
QCD must have domain walls.  Across the wall, the \U1A phase of the
color-superconducting condensate varies from 0 to $2\pi$. Thus, the
same ground state is on both sides of the domain wall and,
consequently, the domain wall is metastable. 
Our proof of the existence and the long lifetime of such
domain walls relies on the following three facts:
(i) the instanton density is small at large chemical potential, 
suppressing the effect of chiral anomaly, 
(ii) the \U1A symmetry is
spontaneously broken, and (iii) the decay constant of the pseudoscalar
singlet boson is large. All these effects are known, from earlier studies,
to occur in the color-superconducting phases of QCD.  However, as far
as we know, their implication for the domain walls has not been
explored. These
domain walls are similar to the walls of Ref.\ \cite{Zhitnitsky}, which
may exist, but no definite statement can be made in this case.
Asymptotic freedom of QCD allows us to assert the existence of 
domain walls in
the high baryon density regime reliably.  The properties of the walls
can be determined by controllable weak-coupling calculations.

We shall also make a rough estimate of the critical chemical potential
above which such domain walls must appear within weak-coupling instanton
calculation. 
This chemical potential is relatively high (though not unreasonably
high).  In the model of QCD with two light flavors, the critical
chemical potential is estimated to be about $6\Lambda_{\rm QCD}\sim 1$ GeV,
close to the scale at which instanton interactions become relevant \cite{SVZ}.
 We
also calculate the semiclassical lifetime of the domain wall and find
it to be exponentially long.

{\em Domain walls in two-flavor high-density QCD.}---%
The simplest model with high-density domain walls is 
QCD with $N_f=2$ massless quark flavors ($u$ and $d$).  This model
is a rather good approximation to realistic quark matter at moderate
densities, such as in the neutron star interiors.  
We recall that the ground state of this model at high
baryon densities is the two-flavor color-superconducting state
\cite{colorSC,2SC}, characterized by the condensation of
diquark Cooper pairs.  These pairs are
antisymmetric in spin ($\alpha,\beta$), flavor ($i,j$) and color ($a,b$) 
indices:
\begin{eqnarray}
  \<q^{ia}_{L\alpha} q^{jb}_{L\beta} \>^* &=&  
  \epsilon_{\alpha\beta} \epsilon^{ij}\epsilon^{abc} X^c \, ,
  \nonumber \\
  \<q^{ia}_{R\alpha} q^{jb}_{R\beta} \>^* &=&  
  \epsilon_{\alpha\beta} \epsilon^{ij}\epsilon^{abc}Y^c \, .
  \label{2flavor}
\end{eqnarray}
The condensates $X^c$ and $Y^c$
are complex color three-vectors.  In the ground state, $X^c$ and $Y^c$ are aligned
along the same direction in the color space, and they break
the color SU(3)$_c$ group down to SU(2)$_c$.  The lengths
of these vectors are equal $|X|=|Y|$ and can be computed
perturbatively (see below).

In perturbation theory, there is a degeneracy of the ground state
with respect to 
the relative U(1) phase between $X^a$ and $Y^a$.  This is  
due to the \U1A symmetry of the QCD Lagrangian at the classical level.
This fact implies that the \U1A symmetry is spontaneously broken
by the color-superconducting condensate.  Since this is a global symmetry, 
its breaking gives rise to a Goldstone boson, which we
denote by $\eta$ since it carries the same quantum numbers as the
$\eta$ boson in vacuum.

It is possible to construct the field corresponding to
$\eta$ boson explicitly. The following object,
\begin{equation}
  \Sigma=XY^\+\equiv X^aY^{a*} \, ,
  \label{SigmaXY}
\end{equation}
in contrast to $X$ and $Y$, is a gauge-invariant order
parameter.
Furthermore $\Sigma$ carries a
nonzero \U1A charge.  Indeed, under the \U1A
rotations 
\begin{equation}
  q\to e^{i\gamma_5\alpha/2}q \, ,
  \label{qalpha}
\end{equation}
 the fields (\ref{2flavor})
transform as $X\to e^{-i\alpha}X$, $Y\to e^{i\alpha}Y$, and therefore
$\Sigma\to e^{-2i\alpha}\Sigma$.  Thus, the color-superconducting
ground state, in which $\<\Sigma\>\neq0$, breaks the \U1A symmetry.  
The Goldstone mode $\eta$ of this symmetry breaking is described by
the phase $\varphi$ of $\Sigma$,
\begin{equation}
  \Sigma=|\Sigma|e^{-i\varphi} \, . 
\end{equation}
Under the \U1A rotation (\ref{qalpha}), $\varphi$ transforms as
\begin{equation}
  \varphi \to \varphi + 2\alpha \, .
  \label{phi_transf}
\end{equation}

At low energies, the dynamics of the Goldstone mode $\varphi$ is
described by an effective Lagrangian, which, to leading order in
derivatives, must take the following form,
\begin{equation}
  L = f^2 [(\d_0\varphi)^2 - u^2 (\d_i\varphi)^2] \, .
  \label{Leffnomass}
\end{equation}
This Lagrangian (\ref{Leffnomass}) contains two free parameters: the
decay constant $f$ of the $\eta$ boson, and its velocity $u$.  In
general, the velocity $u$ of the $\eta$ boson may be different from
the speed of light (i.e., unity) since the Lorentz invariance is
violated by the dense medium. For large
chemical potentials $\mu\gg\LambdaQCD$, the 
leading perturbative values for $f$
and $u$ have been determined by
Beane, Bedaque and Savage \cite{BBS}:
\begin{equation}
  f^2 = {\mu^2\over 8\pi^2} \, , \qquad  u^2 = {1\over3} \, .
\end{equation}
In particular, the velocity of
the $\eta$ bosons, to this order, is equal
to the speed of sound.  The fact that 
$f\sim\mu$ 
plays an important role in our further discussion.

It is well known that the \U1A symmetry is not a true symmetry of
the quantum theory, even when quarks are massless.  The violation of
the \U1A symmetry is due to nonperturbative effects of
instantons.  Since at large chemical potentials the instanton density
is suppressed (see below), the $\eta$ boson still exists but acquires 
a finite mass.  In other words, the anomaly adds a potential energy 
term $V_{\rm inst}(\varphi)$ to the Lagrangian
(\ref{Leffnomass}),
\begin{equation}
  L = f^2 [(\d_0\varphi)^2 - u^2 (\d_i\varphi)^2] -
      V_{\rm inst}(\varphi) \, .
  \label{LeffV}
\end{equation}
The curvature of $V_{\rm inst}$ around $\varphi=0$ determines the mass
of the $\eta$.

A standard symmetry argument determines periodicity of 
$V_{\rm inst}(\varphi)$.
One can formally restore the \U1A symmetry  by
accompanying (\ref{qalpha}) by a rotation of the $\theta$-parameter
\begin{equation}
  \theta \to \theta + N_f \alpha = \theta + 2\alpha \, .
  \label{theta_transf}
\end{equation}
This symmetry must be preserved in the effective Lagrangian, so the
latter is invariant under (\ref{phi_transf}) and (\ref{theta_transf}).
This means that the potential $V_{\rm inst}$ is a function of the
variable $\varphi-\theta$, unchanged under \U1A.
Since we know that the physics is periodic in $\theta$ with
period $2\pi$, we can conclude that, at the physical value of the
theta angle $\theta=0$, $V_{\rm inst}$ is a periodic function of
$\varphi$ with period $2\pi$. 

Moreover, at large $\mu$, $V_{\rm inst}$ can be found from instanton
calculations explicitly.  
The infrared problem that plagues these calculations in
vacuum disappears at large $\mu$: 
large instantons are
suppressed due to
Debye screening \cite{GPY,Shuryak_mu}. 
As a result, most instantons have small size 
$\rho\sim{\cal O}(\mu^{-1})$ 
and the dilute instanton gas approximation
becomes reliable. One-instanton contribution, proportional to 
$\cos(\varphi-\theta)$, dominates in $V_{\rm inst}$. Therefore,
\begin{equation}
  V_{\rm inst}(\varphi) = -a \mu^2\Delta^2\cos\varphi \, ,
  \label{Vinst}
\end{equation}
where $\Delta$ is the Bardeen-Cooper-Schrieffer (BCS) gap, and
$a$ is a dimensionless function of $\mu$ which will be found
later.  Here we only note that $a$ vanishes in the limit
$\mu\to\infty$.  This is an important fact, since it implies
that the mass of the $\eta$ boson,
\begin{equation}
   m = \sqrt{a\over2} \, {\mu\over f} \Delta 
   = 2\pi \sqrt a \Delta \, ,
   \label{meta}
\end{equation}
becomes much smaller than the gap $\Delta$ at large $\mu$.
In this case the effective theory
(\ref{LeffV}) is reliable, since meson modes other than $\eta$
have energy of order $\Delta$, i.e., are much heavier than $\eta$ and
decouple from the dynamics of the latter.

The Lagrangian (\ref{LeffV}) with the potential (\ref{Vinst}) is just
the sine-Gordon model, in which there exist domain-wall solutions to
the classical equations of motion.  The profile of the wall parallel
to the $xy$ plane is
\begin{equation}
   \varphi = 4 \arctan e^{mz/u} \, ,
\end{equation}
so the wall interpolates between $\varphi=0$ at $z=-\infty$ and
$\varphi=2\pi$ at $z=\infty$.  The tension of the domain wall is
\begin{equation}
  \sigma = 8\sqrt{2a}\, uf\mu\Delta  
  \, .
  \label{sigma}
\end{equation}
A good analog of this domain wall is the $N=1$ axion domain wall,
which also interpolates between the same vacuum.

{\em Decay of the domain wall.}---%
It is important to understand the mechanism of the decay of the wall.
It has nothing to do with the decay of $\eta$ meson quanta, 
which are due to $\eta$ coupling to photons, ungapped 
quarks, or the gluons of the unbroken SU(2)$_c$ subgroup. 
The domain
wall is already a local minimum of the energy, and the decay
of its excitations means only that the fluctuations 
around this minimum, corresponding to deformations
of the wall, are damped. 

The domain wall is not stable because the same ground state
is on both of its sides: $\varphi=0$ and $\varphi=2\pi$ are equivalent.
The  instability is due to higher energy meson modes integrated
out and not present in the Lagrangian (\ref{LeffV}). One can
visualize the effect of these modes by considering an effective
potential which depends on the magnitude $|\Sigma|$ as well as
on the phase $\varphi$ of the order parameter $\Sigma$.
This potential has the shape of a Mexican hat,  slightly tilted by an
angle proportional to $a$.  A similar picture is discussed 
in Ref.\ \cite{Zhitnitsky}, except that
the tilt of the hat is very small in our case.
The domain wall is a configuration
that, as a function of the coordinate perpendicular to the wall,
starts from the global minimum, goes along the
valley, and returns to the starting point. One can continuously
deform this configuration into a trivial constant one by pulling
the looplike trajectory over the top of the hat.

As in the case of the axion wall \cite{axions,Kibble}, 
this deformation has to be
done in a finite area of the wall first, thus creating a hole.
If this hole exceeds the critical size, it will expand, destroying
the wall. On the rim of the hole the magnitude of $|\Sigma|$
vanishes. The field configuration around the rim is a vortex: on a closed
path around the rim, $\varphi$ changes by $2\pi$.  
The decay of
the wall is a quantum tunneling process in which a hole bounded by a
closed circular vortex line is nucleated. The
semiclassical probability of this process is
\begin{equation}
  \Gamma \sim  \exp\biggl( -{16\pi\over3}{\nu^3\over u\sigma^2} \biggr)
  \, ,
  \label{hole}
\end{equation}
where $\nu$ is the tension of the vortex line in the limit of massless
boson, $m=0$.  The factor $1/u$ in the exponent of Eq.\
(\ref{hole}) is due to the fact that $u$ plays the role of light speed
for the effective dynamics of the Goldstone boson [see Eq.\
(\ref{Leffnomass})].

To find $\Gamma$ we still need to compute the vortex tension $\nu$.
Since the vortex is a global string, its tension is logarithmically
divergent,
\begin{equation}
  \nu = 2\pi u^2 f^2 \ln {R\over R_{\rm core}}
      = 2\pi u^2 f^2 \ln (R\Delta) \, ,
  \label{nu}
\end{equation}
where $R$ is
a long-distance cutoff to be specified later,
and $R_{\rm core}$ is the
size of the core of the vortex line, which is the short-distance cutoff.
$R_{\rm core}\sim 1/\Delta$ since
$\Delta$ is the momentum scale at which the effective Lagrangian
description breaks down.  We are helped by the fact that the vortex
tension is dominated by the region outside the core, so the effective
Lagrangian (\ref{Leffnomass}) is sufficient for computing $\nu$ to the
logarithmic accuracy.
By using Eqs.\ (\ref{sigma}) and (\ref{nu}), and taking $R$ to be 
the thickness of the wall,
we find the decay rate to be
\begin{equation}
  \Gamma \sim \exp\biggl(-{\pi^4\over3}{u^3\over a}{f^4\over\mu^2\Delta^2}
  \ln^3{1\over\sqrt{a}}
  \biggr) \, .
  \label{Gamma}
\end{equation}
Since $f\sim\mu\gg\Delta$, and $a$ decreases with increasing $\mu$, the
decay rate is exponentially suppressed at high $\mu$.  Thus, we have
shown that (i) domain walls exist in the limit of large chemical
potentials, and (ii) they are metastable with parametrically long
lifetime. These conclusions are valid in the regime of very 
large chemical potentials $\mu$, where our calculations are under control.

{\em Calculation of the potential.}---%
What happens at smaller $\mu$? The most interesting possibility
is that the walls persist down to $\mu=0$ as advocated in 
Ref.\ \cite{Zhitnitsky}
using large-$N_c$ arguments. Another possibility is that,
as the description based on the Lagrangian (\ref{LeffV}) breaks
down, the walls disappear. This happens when the mass of the
$\eta$ excitation becomes comparable to $2\Delta$, the
typical energy scale for higher mesons \cite{Shovkovy}.
From Eq.\
(\ref{meta}), one derives the following condition when our effective
Lagrangian description is under control:
\begin{equation}
  a(\mu) \lesssim 1/\pi^2 \, .
  \label{criterion}
\end{equation}
We shall now evaluate the function $a(\mu)$.

To compute $V_{\rm inst}(\varphi)$, we start from the
instanton-induced effective four-fermion interaction 
\cite{tHooft,SVZ,SchaeferShuryak},
\begin{eqnarray}
   L_{\rm inst}&=& \int\!d\rho\, n_0(\rho) 
  \biggl({4\over3}\pi^2\rho^3\biggr)^2 \biggl\{
  (\bar u_R u_L)(\bar d_R d_L) + \nonumber \\
  &+& {3\over32} \biggl[ (\bar u_R\lambda^a u_L)(\bar d_R\lambda^a d_L)
\nonumber\\
  &-& {3\over4}(\bar u_R\sigma_{\mu\nu}\lambda^a u_L)
    (\bar d_R\sigma_{\mu\nu}\lambda^a d_L) \biggr]
  \biggr\}  + {\rm H.c.}
  \label{inst_vertex}
\end{eqnarray}
By taking the average of Eq.\ (\ref{inst_vertex}) over the
superconducting state (\ref{2flavor}), one finds $V_{\rm inst}$, and
confirms that it is proportional to $\cos\varphi$ as in Eq.\ (\ref{Vinst}). 
In the ground state,
\begin{equation}
  |X| = |Y| = {1\over2} \int\!{d^4p\over(2\pi)^4}\,
  {\Delta(p_0) \over p_0^2 + (|{\bf p}|-\mu)^2 + \Delta^2(p_0)} \, ,
\end{equation}
where $\Delta(p_0)$ is the momentum-dependent BCS gap.  Using the
perturbative result \cite{Son:1999uk}, 
\begin{equation}
  \Delta(p_0) = \Delta \cos\biggl( {g\over 3\sqrt{2}\pi}
  \ln{p_0\over\Delta} \biggr) , \qquad
  \Delta\lesssim p_0\lesssim\mu
  \, ,
\end{equation}
we find
\begin{equation}
  |X| = {3\over 2\sqrt{2}\pi} {\mu^2\Delta\over g} \, .
\end{equation}
Averaging Eq.\ (\ref{inst_vertex}) in the superconducting background, we
find, after some calculations
\begin{equation}
  V_{\rm inst}(\varphi) = -\int\!d\rho\, n_0(\rho)
  \biggl({4\over3}\pi^2\rho^3\biggr)^2 12|X|^2\cos\varphi \, .
\label{Vinst2}
\end{equation}
Using the standard formula for the instanton density at finite chemical
potential \cite{Shuryak_mu,SchaeferShuryak}
\begin{equation}
  n_0(\rho) = C_N \biggl({8\pi^2\over g^2}\biggr)^{2N_c} \rho^{-5}
  \exp\biggl(-{8\pi^2\over g^2(\rho)}\biggr) e^{-N_f\mu^2\rho^2} 
\end{equation}
with
\begin{equation}
  C_N = {0.466 e^{-1.679N_c} 1.34^{N_f}\over(N_c-1)!(N_c-2)!} \, ,
\end{equation}
we arrive at the final result
\begin{equation}
  a = 5 \times 10^4 \biggl(\ln{\mu\over\LambdaQCD}\biggr)^7
  \biggl({\LambdaQCD\over\mu}\biggr)^{29/3} \, .
  \label{amu}
\end{equation}
Thus $a\to0$ when $\mu\to\infty$, so at sufficiently large $\mu$ the
criterion (\ref{criterion}) is satisfied.  However, due to the 
large numerical constant in Eq.\
(\ref{amu}) the critical $\mu$ is quite high: $\mu_{\rm
crit}\sim6\LambdaQCD\sim1$ GeV.  This result should be taken with some care
due to the uncertainty in the criterion (\ref{criterion}).  However,
since $a$ depends quite sensitively on $\mu$, 
it is reasonable to expect that our
estimate is not far from the true value.

In the estimate above we neglected the contribution from large
instantons, which arise from the unbroken SU(2)$_c$ sector of the
theory.  This sector is governed by a pure Yang-Mills theory
with the confinement scale $\LambdaQCD'\sim\Delta
\exp[-\mbox{const}\cdot\mu/(g\Delta)]$
\cite{gcm}. The nonperturbative contribution of large SU(2)$_c$ instantons
is of order $(\LambdaQCD')^4$. Since $\LambdaQCD'$ is exponentially
small, this contribution is negligible compared to
that from small SU(3)$_c$ instantons.

Inclusion of quark masses does not change the domain walls in a
substantial way.  
The mass contribution to the potential has been found in Ref.\
\cite{BBS}, 
\begin{equation}
  V_{\rm mass} = - b\, m_u m_d\, \Delta^2 \cos\varphi \, ,
  \label{Vmass}
\end{equation}
where $b\sim1$.  Eq.\
(\ref{Vmass}) has the same $\varphi$ dependence as $V_{\rm inst}$,
therefore, in all previous formulas one should replace $a$ by $a+b\, m_u
m_d/\mu^2$.

{\em Discussion.}---%
It would be interesting to investigate possible astrophysical
consequences of the high-density QCD walls. In particular, one
would like to know if such walls can be created inside neutron
stars. To describe the motion of the wall one may need more
than just the effective Lagrangian (\ref{LeffV}): the coupling of $\eta$
to ungapped quarks and SU(2)$_c$ gluons could be important. 
The moving wall 
may radiate quark-hole pairs, gluons, or photons, slowing down the collapse
of a closed domain wall surface.  

It is possible to generalize our results to the color-flavor-locking 
state of $N_f=3$ QCD \cite{CFL}.  The \U1A
symmetry is also spontaneously broken in this case.  The role of the
$\eta$ boson is played by the $\eta'$ meson, which is also light
at high densities \cite{inverse}. The instanton-induced $\eta'$ 
potential has a form similar to (\ref{Vinst}) \cite{MT}:
\begin{equation}
  V_{\rm inst}(\varphi) = -a'\cdot (m_s/\mu)\ 
  \mu^2\Delta^2\cos\varphi \, ,
\end{equation}
where the evaluation of dimensionless function $a'$ is very much the same
as our calculation of $a$ and amounts to inserting extra factor
$m_s\rho$ into (\ref{Vinst2}).
Thus one expects the domain walls to 
exist and to be metastable at large $\mu$.  
Due to the mixing between the neutral mesons \cite{inverse}, 
the $\pi^0$ and $\eta$ fields are also nontrivial on the wall.
The domain walls also exist in QCD with
large isospin density \cite{isospin}. This case is interesting,
since it can be studied by a Monte Carlo lattice simulation.

We are indebted to G.~Gabadadze, L.~McLerran, R.D.~Pisarski,
T.~Sch\"afer, M.A.~Shifman, I.A.~Shovkovy, E.V.~Shuryak, A.~Vainshtein, and
M.B.~Voloshin for discussions.  We thank RIKEN, Brookhaven National
Laboratory, and U.S.\ Department of Energy 
for providing the facilities essential for the completion of this work.
D.T.S. is supported, in part, by
a DOE OJI grant.  A.R.Z. is supported, in part, by NSERC of
Canada.


\begin{thebibliography}{99}
\vspace{-0.5cm}
\bibitem{Rajaraman} R.~Rajaraman, {\em Solitons and Instantons}
(North-Holland, Amsterdam, 1982).

\bibitem{axions}
A.~Vilenkin and E.P.S.~Shellard, 
{\em Cosmic strings and other topological defects}
(Cambridge University Press, Cambridge, UK, 1994).

\bibitem{Zhitnitsky}
M.M.~Forbes and A.R.~Zhitnitsky,
hep-ph/0008315.

\bibitem{GabadadzeShifman}
G.~Gabadadze and M.~Shifman,
Phys. Rev. D {\bf 62}, 114003 (2000).

\bibitem{KPT}
D.~Kharzeev, R.D.~Pisarski, and M.H.~Tytgat,
Phys.\ Rev.\ Lett.\  {\bf 81}, 512 (1998);
T.~Fugleberg, I.~Halperin, and A.~Zhitnitsky,
Phys.\ Rev.\  D {\bf 59}, 074023 (1999).


\bibitem{SVZ}
M.A.~Shifman, A.I.~Vainshtein, and V.I.~Zakharov,
Nucl.\ Phys.\  {\bf B163}, 46 (1980);
{\bf B165}, 45 (1980).

\bibitem{colorSC}
        B.C. Barrois, 
Ph.D. thesis, California Institute of Technology, 1979;
        D. Bailin and A. Love, Phys. Rep. {\bf 107}, 325 (1984).

\bibitem{2SC}
        M. Alford, K. Rajagopal, and F. Wilczek,
               Phys. Lett. B {\bf 422}, 247 (1998);
        R. Rapp, T. Sch\"afer, E.V. Shuryak, and M. Velkovsky,
                Phys. Rev. Lett. {\bf 81}, 53 (1998). 

\bibitem{BBS}
S.R.~Beane, P.F.~Bedaque, and M.J.~Savage,
Phys.\ Lett.\  B {\bf 483}, 131 (2000).


\bibitem{GPY}
D.J.~Gross, R.D.~Pisarski, and L.G.~Yaffe,
Rev.\ Mod.\ Phys.\  {\bf 53}, 43 (1981).

\bibitem{Shuryak_mu}
E.V.~Shuryak, Nucl. Phys. {\bf B203}, 140 (1982).

\bibitem{Kibble}
T.W.~Kibble, G.~Lazarides, and Q.~Shafi,
Phys.\ Rev.\ D  {\bf 26}, 435 (1982).

\bibitem{Shovkovy}
V.A.~Miransky, I.A.~Shovkovy, and L.C.~Wijewardhana,
Phys.\ Rev.\  D {\bf 62}, 085025 (2000).

\bibitem{tHooft}
G.~'t Hooft,
Phys.\ Rev.\  D {\bf 14}, 3432 (1976).


\bibitem{SchaeferShuryak}
T.~Sch\"afer and E.V.~Shuryak,
Rev.\ Mod.\ Phys.\  {\bf 70}, 323 (1998).

\bibitem{Son:1999uk}
D.T.~Son,
Phys.\ Rev.\ D {\bf 59}, 094019 (1999).


\bibitem{gcm}
D.H.~Rischke, D.T.~Son, and M.A.~Stephanov,
hep-ph/0011379.

\bibitem{CFL}
M.~Alford, K.~Rajagopal, and F.~Wilczek,
Nucl.\ Phys.\  {\bf B537}, 443 (1999).

\bibitem{inverse}
D.T.~Son and M.A.~Stephanov,
Phys.\ Rev.\ D {\bf 61}, 074012 (2000); 
Erratum: {\bf 62}, 059902 (2000).

\bibitem{MT}
C.~Manuel and M.H.~Tytgat,
Phys.\ Lett.\  B {\bf 479}, 190 (2000).

\bibitem{isospin}
D.T.~Son and M.A.~Stephanov,
Phys. Rev. Lett. {\bf 86}, 592 (2001);
hep-ph/0011365.



\end{thebibliography}
\end{document}